\documentclass[fleqn,twoside]{article}
\usepackage{espcrc2}

\usepackage{graphicx}
\newcommand{\AmS}{{\protect\the\textfont2
  A\kern-.1667em\lower.5ex\hbox{M}\kern-.125emS}}

\title{Massive Neutrinos in Cosmology}

\author{M. Fukugita\address[ICRR]{Institute for Cosmic Ray Research, 
        University of Tokyo, \\ 
        Kashiwa 277 8582, Japan}}

       \def\lsim{\mathrel{\mathpalette\vereq<}}
\def\gsim{\mathrel{\mathpalette\vereq>}}
\catcode`\@=11
\def\vereq#1#2{\lower3pt\vbox{\baselineskip1.5pt \lineskip1.5pt
\ialign{$\m@th#1\hfill##\hfil$\crcr#2\crcr\sim\crcr}}}
\catcode`\@=12
\begin{document}

\begin{abstract}
The roles of massive neutrinos in cosmology --- in leptogenesis 
and in the evolution of mass density fluctuations --- are reviewed.
Emphasis is given to the limit on neutrino mass from these
considerations.   

\vspace{1pc}
\end{abstract}

\maketitle

\section{INTRODUCTION}

There are two places in cosmology where the mass of neutrinos would
play a significant role. One is 
leptogenesis in the early universe, and the other is the evolution
of mass density fluctuations that are explored with cosmic
microwave background (CMB) fluctuations and the formation
of cosmic structure. There is one more place where 
the presence of neutrinos is generally important --- primordial 
nucleosynthesis, but the effect of neutrino mass is negligible
unless it is unrealistically heavy.

In this talk I discuss at some length the effect of the neutrino mass on 
the CMB fluctuations and cosmic structure formation,
and the limit on the neutrino mass therefrom: 
we saw a substantial progress in our understanding after 
the reports of {\it Wilkinson Microwave Anisotropy Probe} (WMAP) 
\cite{wmap} and
{\it Sloan Digital Sky Survey} (SDSS) \cite{york} 
but also some confusions still
remain. We start with a brief mention about the status of leptogenesis.

\section{LEPTOGENESIS}

One of the promising ideas for baryogenesis is the generation of
baryon asymmetry via leptogenesis from the Majorana mass term in the
presence of the action of sphalerons of electroweak interactions \cite{FY86}. 
The simplest scenario assumes delayed decay of a thermally produced
heavy Majorana neutrino to a scalar particle and a light neutrino, 
and its conjugate. For the delayed decay scenario to
work there is a lower limit on the smallest heavy Majorana
neutrino mass, as known from the argument of GUT baryogenesis 
\cite{toussaint,weinberg,yoshimura}, which in turn leads
to an upper limit on the mass of light left-handed neutrinos
via the sea-saw mechanism. 
The latest analyses with the Boltzmann equation 
yield \cite{BDiBP03,Giudice,BDiBP05}
\begin{equation}
m_{\nu_i}<0.1 {\rm eV},
\end{equation}
for all species of neutrinos.

It is also desirable to impose the condition that pre-existing
baryon asymmetry that may have arisen from some baryon or lepton 
number violating processes at, e.g., the Planck mass scale, if any,
is erased so that the prediction of leptogenesis does not
depend on the initial condition. This demands that the 
Majorana neutrino be smaller than a certain value, so that 
lepton number violation takes place fast enough compared with 
the expansion of the Universe. This gives a lower
limit on some effective light neutrino mass \cite{BDiBP03}: 
\begin{equation}
0.001 {\rm eV}<m_{\nu_*}.
\end{equation}
The combination of the Yukawa terms that appear in lepton number violation
does not give the mass term that is written in terms of light neutrino
masses; hence the interpretation of this lower limit needs
some care. This limit does not mean that the lightest neutrino
must satisfy this, but it suggests that the lepton number violation
process is unlikely to erase pre-existing lepton asymmetry if
all neutrino masses are as light as those violating this limit.

There is a recent focus of interest if
this thermal leptogenesis scenario is viable in the world with
supersymmetry, with which the reheating temperature cannot be sufficiently
high to produce needed heavy Majorana neutrinos without
overproduction of gravitinos. For a recent review, see \cite{bpy05}.

Experimental verifications of the leptogenesis scenario need: 

\noindent
(1) neutrinos are of the Majorana type;

\noindent
(2) find some evidence for the presence of the unification 
scale that is relevant to massive Majorana neutrinos
($M\gsim 10^9$ GeV). More specifically, this is the scale where
rank of the unifying group steps down to 4 at which extra U(1) gauge
group is broken (see \cite{fy03} \S 9.3.3 and \S 9.6.1
for a detailed argument).

\noindent
(3) CP violation.  However, it would be even more a surprise 
if CP is conserved (i.e., mass matrix is real) for some reasons.

\section{COSMIC STRUCTURE FORMATION AND MASSIVE NEUTRINOS}

By now we believe we understand the evolution of the universe 
as a whole and of cosmic structure at large scales.  
The latest most important step
to modern cosmology was the discovery of 
fluctuations in the cosmic microwave background by the COBE 
satellite in 1991. This indicated that we are on the right
track to understand the cosmic structure formation. At the
same time it gave compelling evidence for cold
dark matter (i.e., dark matter that was non-relativistic when it 
was decoupled from the thermal bath) that dominates 
the matter component of the Universe. 
In the last ten years the cosmological paradigm was also shifted.
The cosmological constant was an anathema in 1990, but now
all observations converged to pointing towards 
the existence of the cosmological constant that 
dominates the energy of the Universe. It is surprising
that no observations have yielded evidence against the 
$\Lambda$ dominated cold dark matter ($\Lambda$CDM) universe.
The recent results from WMAP and SDSS
strongly supported this standard view based on the 
$\Lambda$CDM Universe, as demonstrated by the  
convergence of the 
cosmological parameters \cite{spergel,tegmark2,eisenstein} to 
\begin{eqnarray}
&&H_0=71\pm 5 {\rm km~s^{-1}Mpc^{-1}},\cr
&&\Omega_m=0.28\pm 0.03, \hskip7mm  \Omega_\Lambda=0.72\pm 0.03,\cr
&&\Omega_m+ \Omega_\Lambda=1.01\pm 0.01\ .
\end{eqnarray}
We also know that baryons amount to only 1/6 of $\Omega_m$, and only 6\%
of them are comprised in stars and stellar remnants.

\begin{figure}
\includegraphics[width=8cm]{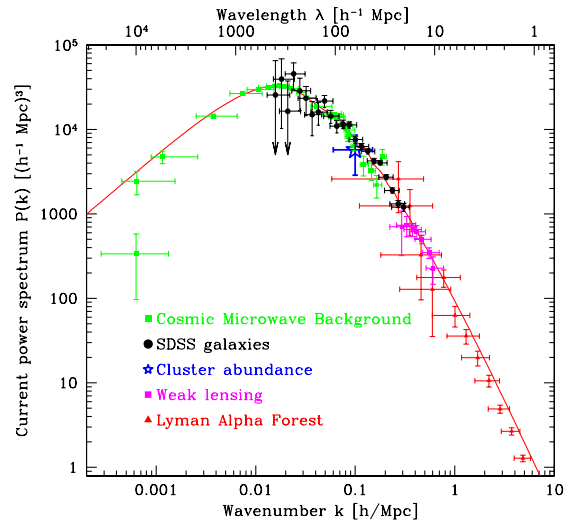}
\caption{Observational power spectrum from CMB fluctuations, galaxy
clustering, gravitational lensing shear, cluster abundances and
Lyman $\alpha$ clouds, compared with the prediction of
$\Lambda$CDM model. The figure is taken from the SDSS paper, Tegmark et al.
\cite{tegmark1}}
\end{figure}

Another important measure is the power spectrum that characterises 
matter fluctuations,
\begin{equation}
P(k)=\int d^3x e^{ik\cdot x}\langle \delta(x)\delta(0)\rangle,
\end{equation}
where $\delta(x)$ is the density contrast at the position $x$.
A variety of observations at vastly different cosmological epochs  
(CMB at $z=1090$, galaxy clustering at $z\sim 0.1-0.4$, gravitational
lensing at $z\sim 0.5-1$, cluster abundances at $z\sim 0.1-0.2$),
when scaled to $z=0$,  yield 
$P(k)$ that is described by $|k|^n T(k)$ with $n=1\pm0.02$
and the transfer function $T(k)$ predicted by the
$\Lambda$CDM model with the cosmological parameters 
specified above \cite{tegmark1,cole} (see Figure 1). 
This consistent
description of $P(k)$ constitutes additional evidence 
that supports the standard model,
nearly scale-invariant initial adiabatic perturbations 
($\delta=3\delta T/T$) growing by gravitational instability in the
$\Lambda$CDM universe. We must add a remark that
no models are known, other than
inflation, that generate these fluctuations consistent with
observations, while 
a successful model of inflation from the particle physics point of view  
is yet to be found.

The massive neutrinos contribute to the mass density of the Universe
by the amount
\begin{equation}
\Omega_\nu=\frac{\sum m_\nu}{94.1~{\rm eV}}h^{-2},
\end{equation}
where $h=0.71$ is the conventional notation for the Hubble constant.
In the following discussion we assume three species of light neutrinos
and do not consider exotic possibilities which are occasionally
discussed by particle physicists.
The neutrino mass and its limits discussed in what follows are of
the order of 1 eV, which is much larger than than 0.05 eV
derived for the mass difference for the limiting case of
the hierarchical-mass neutrinos. 
It is therefore appropriate to
consider three degenerate neutrinos, and we assume this in
our considerations. 

\subsection{Effects of massive neutrinos on the evolution of cosmic
fluctuations}

A successful model is obtained for cosmic structure formation
without having massive neutrinos. This means that massive neutrinos
only disturb the agreement between theory and observations, hence
leading to a limit on the neutrino mass.

The well-known effect of massive neutrinos is relativistic free streaming 
that damps fluctuations within the horizon scale. 
One electron volt neutrinos are still relativistic at 
matter-radiation equality, which takes place at $T\approx 1$ eV,
and then tend to smear fluctuations up to $\sim$100 Mpc comoving scale,
thus diminishing the power of $P(k)$ for these scales. 
This effect becomes stronger as the cosmological mass density of neutrinos,
hence the neutrino mass,  increases (see Figure 2) (see \cite{HET,VKN}).
Therefore, the empirical knowledge of $P(k)$ across large to small
scales gives a constraint on the summed mass of neutrinos.

\begin{figure}
\includegraphics[width=8cm]{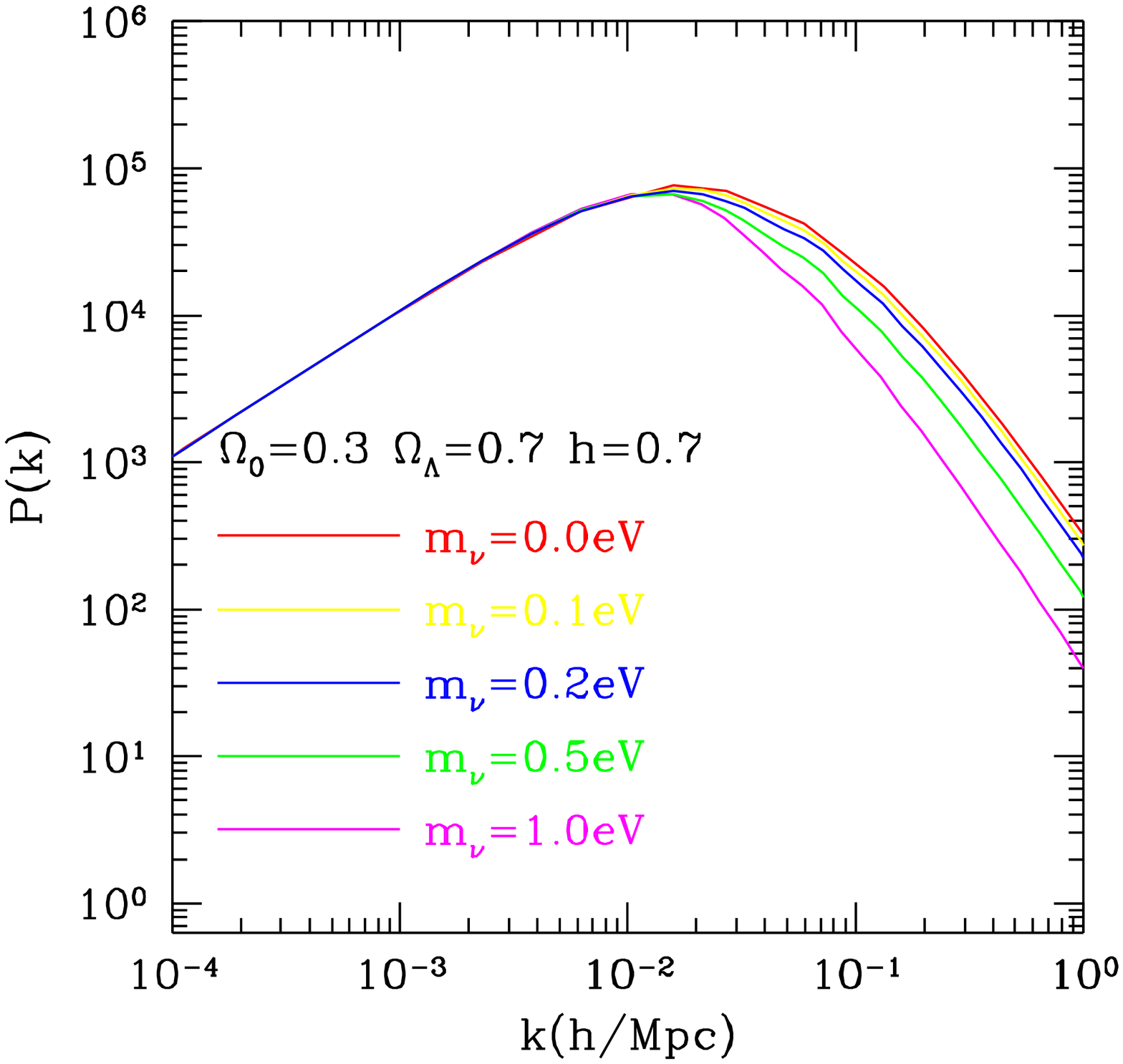}
\caption{Power spectrum predicted in the $\Lambda$CDM model including
massive neutrinos with a degenerate mass specified in the legend. The
top curve represents the case of massless neutrinos}
\end{figure}

The best determination of $P(k)$ for large scales is given 
by CMB multipoles observed
by WMAP \cite{wmap}. 
There are a number of ways to obtain $P(k)$ at small scales:
the use of galaxy clustering,  the cluster abundance,
the Lyman $\alpha$ cloud absorption optical depth, and  
the cosmic shear due to gravitational lensing.
Limits on the neutrino mass are derived from a single or
combined use of these observations. One may also input
other observations, which serve to restrict various cosmological 
parameters, which may indirectly improve the limit on the neutrino mass
in a multidimensional parameter search. 

\noindent
(1) Limits derived from galaxy clustering varies from $\sum m_\nu<0.6$ eV
\cite{spergel} to 2.1 eV \cite{hannestad} at a 95\% confidence, both
authors using 2dFGRS data (see also \cite{allen})\footnote{
Allen et al. \cite{allen} concluded a finite value for the
neutrino mass at 2$\sigma$. Here we take their upper 2$\sigma$ value as
a limit, which is 1 eV.}. The limit using the SDSS data is 
$\sum m_\nu<1.7$ eV \cite{tegmark2}. 
(See also \cite{barger}, where $\sum m_\nu<0.75$ eV is concluded using
SDSS and 2dFGRS.) While I quoted the limits all
obtained by using only CMB and galaxy clustering here, 
the origin of different upper limits is not clear.  
It is possible that the use of SDSS or 2dFGRS would leads to different
limits, as the fall of  $P(k)$ towards smaller scales is
more gentle with the 2dFGRS, thus giving a stronger constraint on $m_\nu$.
A suspect is the convergence of practical applications of 
the Markov chain Monte Carlo method to
calculate the likelihood, which we discuss briefly in section
\ref{sec:3.3}. 

\noindent
(2) The use of the cluster abundance enables us to 
estimate the mass fluctuations that are the quantity
relevant to us. This was
made in the early work \cite{fls},
but no updates were attempted after WMAP.

\noindent
(3) With the  Lyman $\alpha$ cloud absorption power derived from 
fluctuating optical depths,\footnote{
Lyman $\alpha$ clouds are the objects that cause Lyman $\alpha$
absorption lines with the 
neutral hydrogen column density $10^{13}-10^{17}$ cm$^{-2}$
in the quasar spectrum when they intervene the line of
sight to a quasar. The `clouds' are identified with moderate
overdensities of the hydrogen gas of temperature $\sim10^4$K,
governed by photoionisation and adiabatic cooling. 
}
one can explore $P(k)$ at the
smallest scale \cite{croft,mcdonald}, giving the strongest limit
 $\sum m_\nu<0.42$ eV (95\%) \cite{seljak} (see \cite{croft2} for
the earlier work). 
This result, however, is more model dependent
in the sense that one must invoke simulations to extract $P(k)$ from
observed flux power spectrum, which suffer from significant 
uncertainties associated with modelling and simulations.

\noindent
(4) With gravitational lensing one can directly explore 
mass fluctuations.  
For the moment the statistical error is not sufficiently small, 
but this provides us with a promising method. This may also be used
to set the normalisation of the power spectrum derived from galaxy
clustering.

Elgar\o y and Lahav \cite{el-l2} give a summary of limits obtained 
in the literature.

We note that a decrease of spectral index $n$ (red tilting) 
has an effect similar 
to that of massive neutrinos for small scales. 
Hence, unless the length scale 
explored spans a wide range or $n$ is constrained, $m_\nu$ and
$n$ are \underline{positively} correlated, 
i.e., an `artificial' blue tilting would be cancelled by
an appropriate value of $m_\nu$; this weakens the constraint on 
$m_\nu$. Elgar\o y et al. \cite{el} derived from
2dFGRS power spectrum `alone' $\sum m_\nu<2$ eV, but this is
under the assumption that the spectrum is exactly flat, $n=1$. 
Allowing for $n>1$, a sensible limit is lost. We definitely
need CMB data to derive a limit on the neutrino mass.

\subsection{Cosmological limits on the neutrino mass: caveats}

The methods listed above all suffer from different systematic errors.
What we really need is the mass fluctuation 
$\xi_m=\langle \delta_m(x)\delta_m(0)\rangle$
in the linear regime.
We expect that nonlinearity for the length scale relevant to 
galaxy clustering $>10$ Mpc
is only modest (e.g. \cite{smith}), but 
with the use of galaxy clustering we must assume 
that the galaxy distribution traces
mass allowing for some constant factor, i.e., $\xi_g(x)=b^2\xi_m(x)$,
where $b$ is left usually as a free parameter, representing  ``biasing''
of galaxies relative to mass. This
is probably not a bad approximation at the accuracy that concerns 
us at the present, but for higher accuracy we do not know
how good is this approximation: at least we know that biasing
depends on morphology and luminosity of galaxies (e.g., \cite{zehavi}). 
Hence the sample
used must be homogeneous in this regard. 
The systematic difference in $P(k)$ between the two groups must 
also be resolved. % to go beyond the current 1.7 eV limit.
It should also be noted that errors for the galaxy
clustering data given in the literature may not represent properly
the systematics that are associated with the evaluation of
the window function of the survey data analysis.

The use of the cluster abundance is proposed as a way to estimate
the mass fluctuation taking advantage that the cluster mass 
can be estimated with various
means, from velocity dispersion measurements, X ray data, and 
gravitational lensing. For the moment, however, there are  
non-negligible uncertainties in the cluster mass estimate. 
The disadvantage of this method is that one needs a very large
cluster sample to estimate $P(k)$ for varying scales. 

One can explore mass fluctuations in the smallest scale 
with the use of Lyman $\alpha$ cloud absorption power spectrum; 
thus one can get potentially the information most sensitive
to $m_\nu$.
The problem is that one needs substantial corrections to unfold
the matter power spectrum $P(k)$, which can be done only with
simulations. While a great success of numerical
study was to enable us understand the nature of Lyman $\alpha$ clouds
\cite{lyalpha} (see \cite{rauch} for a review), it is not easy 
at a quantitative level to document 
the systematic errors. 
The simulation always suffers
 from mesh effects, and this is particularly true when baryons 
are included. With the inclusion of baryons one has to
deal with stars and their feedback, which is 
a difficult astrophysical problem. 
Specifically for the case of Lyman $\alpha$
clouds, we must be concerned with uncertainties as to the heating 
rate from ionising radiation and thermal history of clouds. 
The estimate of the mean level of
absorption optical depths is an additional source of 
errors. While much work is currently being done,
the assessment of errors in $P(k)$ 
derived from  Lyman $\alpha$ clouds
is not easy a problem.

In principle the use of cosmic shear is the best for
estimating the mass fluctuation. The nonlinearity correction
is still needed, but it can be done with simulations without
baryons, which are relatively simple. A difficulty is that the
signal is small compared with noise so that one needs a very large
galaxy sample to attain good statistics, and that one needs to know
instrumental distortions of images accurately. 

\subsection{Limits on neutrino mass from CMB alone?}
\label{sec:3.3}

There is a controversy as to whether a sensible limit is derived
on the neutrino mass  from 
CMB multipoles alone. This consideration is meaningful since 
one might be able to derive the most robust limit on the neutrino mass 
free from systematics
that are difficult to control with data on galaxies.  
Tegmark et al. \cite{tegmark2}, however, showed
that no limits are obtained from CMB alone: $\sum m_\nu<9$ eV is
allowed at one sigma. Elgar\o y \& Lahav \cite{el-l1} endorsed
this result, emphasising the need to employ large-scale structure 
data. They also suggest that there is a solution that
reproduces CMB multiples only with CDM and massive neutrinos that
satisfy $\Omega_m+\Omega_\nu=1$ without a 
cosmological constant, i.e., the `old' mixed dark matter (MDM) scenario is
viable (see Figure 3), with the only price being a low Hubble constant. 

\begin{figure}
\includegraphics[width=8cm]{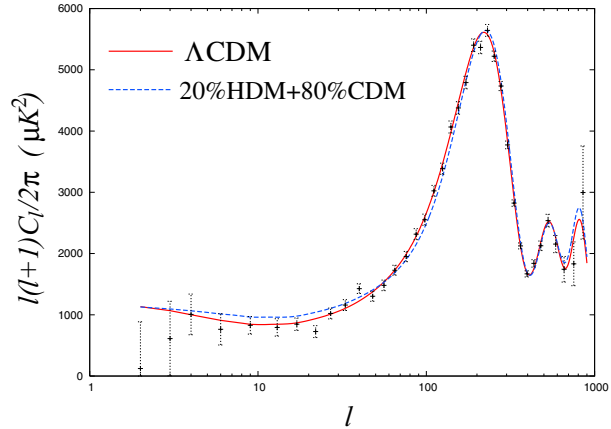}
\caption{CMB multipoles for the MDM like solution suggested by
Elgar\o y \& Lahav ($\sum m_\nu=4$ eV, dotted curve)
are compared with those for the best $\Lambda$CDM solution (with
massless neutrinos, solid curves). The
$\chi^2$ is higher by 50 with this MDM like solution.
The data are from WMAP.}
\end{figure}

On the other hand, Ichikawa, Fukugita \& Kawasaki \cite{IFK05}
claimed that a sensible
limit $\sum m_\nu<2$ eV (at a 95\% confidence) can be derived from CMB
(WMAP data) alone. They argued that the effects of massive neutrinos
on CMB multipoles 
cannot be absorbed into the shift of cosmological parameters (notably the
Hubble constant)
if neutrinos are already nonrelativistic at the
recombination epoch, i.e., $m_\nu\gsim 3T_{\rm rec}\simeq 0.55$ eV.
The principle is that the presence of non-relativistic neutrinos 
causes a decay of
gravitational potential in pre-recombination epoch, which amplifies the
acoustic oscillation that appears as the second and 
third peaks \cite{DGS} (while
the first peak receives little the effect from the potential decay,
since the multipoles corresponding to free streaming of the
$m_\nu\gsim 0.55$ eV neutrino are $\ell\gsim 250$ which are larger than
the position of the first peak, $\ell_1=220$).

The shifts of the position and the height of the first peak, 
dominantly caused by modifications of the angular diameter distance 
and of the integrated Sachs-Wolfe effect, respectively, are still
significant for $m_\nu<0.55$ eV,
but the heights of the second and third
peaks relative to the first are hardly modified.
These effects 
can be absorbed into the shift of the other cosmological parameters.
When the second and third peaks change in addition for  
$m_\nu>0.55$ eV, however, 
one cannot absorb the effects into the shift of cosmological
parameters. 
This is the basic mechanism how the constraint on $m_\nu$ is derived
 from CMB multipoles. 
The Hubble constant prior as claimed in \cite{el-l1} is not essential.
The numerical analysis of Ichikawa et al. showed
that the increase of $\chi^2$ with the neutrino mass is very slow
till $\sum m_\nu\approx 1.5$ eV and gave
%\begin{equation}
$\sum m_{\nu_i}<2.0~{\rm eV}$ 
at a 95\% C.L.
%\end{equation}

Figure 4 compares the CMB multipoles for the case of $\sum m_{\nu_i}=2.0$ eV 
with the best model having massless neutrinos that gives the global $\chi^2$
minimum. This demonstrates the accuracy of the WMAP data,
which has a power to distinguish the two curves.
It is argued \cite{IFK05} 
that one cannot obtain a limit significantly better than 
$0.55\times3\simeq1.6$ eV
even if the precision of CMB multipoles is improved.

\begin{figure}
\includegraphics[width=8cm]{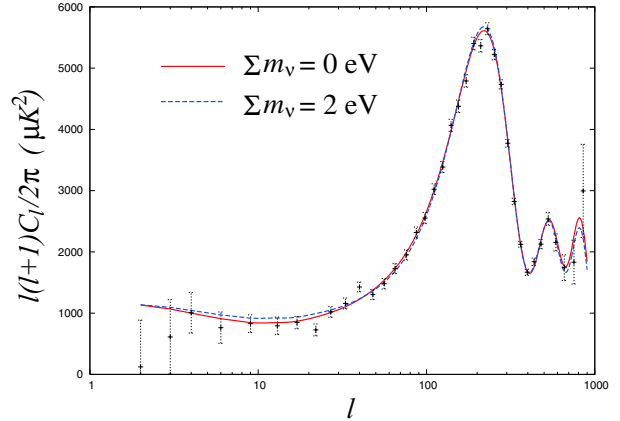}
\caption{CMB multipoles for the $\Lambda$CDM solution with massive neutrinos
that is allowed just at a 95\% confidence ($\sum m_\nu=2.0$ eV, dashed 
curve) are compared with those for the best $\Lambda$CDM solution
with massless neutrinos (solid curve).The data are from WMAP.} 
\end{figure}

Note that an increase of the peaks at higher $\ell$ modes with
massive neutrinos mimics blue tilting, and hence  
there emerges a \underline{negative} correlation 
between $n$ and $m_\nu$, which is opposite to that 
expected from the effect on the power spectrum,
as explained in the previous subsection.

Elgar\o y \& Lahav \cite{el-l2} 
ascribed the discrepancy between Tegmark et al./
Elgar\o y \& Lahav and Ichikawa et al. to the difference of priors,
but they did not give an analysis that supports this statement. 
I would ascribe the origin
to insufficient samplings of the Markov-chain Monte Carlo methods the
former authors adopted to estimate the likelihood functions, especially
away from the point at $m_\nu=0$. When a chain is not sufficiently 
long the Markov-chain Monte Carlo methods often result in likelihood 
functions that differ in shapes depending on the initial condition.
This was the reason why Ichikawa et al. preferred to use a
deterministic algorithm to search for the minima at given $m_\nu$
with successive parabolic approximations, with the further use of the Vegas 
integration code to check for the likelihood function.
In this approach, priors do not play an important role.   
They also remarked that they could not reproduce the
likelihood function of Tegmark et al. as a function of $n$ away from 
the global minimum, indicating that their likelihood function
may probably be incorrectly evaluated, 
whereas the behaviour around the global minimum 
was reproduced very well.

Ichikawa et al. further searched for MDM like solutions, but failed to
find any with $\Delta\chi^2$ smaller than 16. 
%The ``best'' HDM-like solution they obtained is shown in Figure 3 above. 
The curves shown in Figure 3 is the solution suggested by 
Elgar\o y \& Lahav \cite{el-l1} which gives $\chi^2$ higher than the best
solution by $\Delta\chi^2=50$. 
Although the shape of the 
multipoles appears similar to that of the best $\Lambda$CDM solution to eyes,
the accuracy of the WMAP data rules out the MDM model at a high confidence.   

\subsection{Conclusions and prospects}

I discussed that there are mainly two effects that lead to constraints 
on the neutrino mass, damping of the small scale power due to neutrino
free streaming, and decay of the gravitational potential in the 
pre-recombination epoch that amplifies the acoustic oscillation
in high multipole modes.  

The limit on the neutrino mass from cosmic fluctuations
is $\sum m_{\nu_i}<2$ eV (95\%) from CMB alone, which is subject to
systematic errors to the least extent. The assumption is that $\Lambda$CDM
model is correct and the power spectrum obeys a power law, allowing
for a small departure from the power law to the extent
as expected in slow-roll inflation. 
The limit obtained by combining the galaxy clustering
with the CMB data is $\sum m_{\nu_i}\lsim 1-1.7$ eV (95\%).
Unfortunately, the error arising from the Monte Carlo sampling is not well
documented, and 
an accurate limit is yet to be found for a given set
of the clustering data. 
A systematic error in the
power spectrum between two major groups is also yet to be
resolved.
The current most stringent limit is
$\sum m_{\nu_i}<0.42$ eV (95\%) derived with the use of Lyman $\alpha$
cloud absorption power spectrum by unfolding the matter power
spectrum  with the aid of simulations of the
 Lyman $\alpha$ cloud. It is not easy to assess its reliability at the
moment.

The accuracy of the CMB multipole data will be greatly increased, 
especially at the higher multipoles with the launch of {\it Planck}
\cite{planck}. 
However, a straightforward analysis of its data is unlikely to bring a
drastic improvement in the constraint on the neutrino mass.
Ichikawa et al. considered that the limit that can be achieved
 from CMB alone will be $\sum m_{\nu_i}<1.2$ eV. Efforts to
remove systematic errors are needed for small scale clustering 
data. Particularly important is to enhance the reliability of
simulations in a way convincing to everybody and to document errors
involved in the results of simulations. 

Clearly, the goal is to give a constraint as small as 
$\sum m_{\nu_i}\approx 0.05$ eV, which is the minimum mass indicated by
neutrino oscillation experiments.
Kaplinghat et al. \cite{kaplinghat} proposed to use gravitational
lensing signals in CMB polarisation. They forecast that one would get
a limit $\sum m_{\nu_i}<0.15$ eV with {\it Planck}, but need
a new project particularly sensitive to CMB polarisation to get to
 $<0.04$ eV. An accurate simulation is also needed
to unfold non-linear effects of mass clustering in small scales.  
Wang et al. \cite{wang} proposed to use large cluster surveys.
Their forecast is $\sum m_{\nu_i}<0.03$ eV
by combining cluster survey using LSST \cite{lsst} with multipoles from
{\it Planck}. Here the
estimate of cluster mass is crucial, which they hope to perform
using simulations of clusters.

%\section{MISCELLANEOUS TOPICS ON ASTROPHYSICAL NEUTRINOS}

\end{document}